# Imputation and Missing Indicators for handling missing data in the development and implementation of clinical prediction models: a simulation study

Rose Sisk[1], Matthew Sperrin[1,2], Niels Peek[1,2,3], Maarten van Smeden[4], Glen P. Martin[1,2]

1. Division of Informatics, Imaging and Data Science, Faculty of Biology, Medicine and Health, University of Manchester, Manchester Academic Health Science Centre, Manchester, United Kingdom

2. Alan Turing Institute, London, UK

3. NIHR Manchester Biomedical Research Centre, Faculty of Biology, Medicine and Health, University of Manchester, Manchester Academic Health Science Centre, Manchester, UK

4. Julius Center for Health Sciences and Primary Care, University Medical Center Utrecht, Utrecht University, Utrecht, Netherlands

# 1 Abstract

## 1.1 Background

Existing guidelines for handling missing data are generally not consistent with the goals of clinical prediction modelling, where missing data can occur at any stage of the model pipeline; development, validation or application. Multiple imputation (MI), often heralded as the gold standard approach, can be challenging to apply in the clinic. Clearly, the outcome cannot be in the imputation model, as recommended under MI, at prediction time. Regression imputation (RI) is an alternative strategy, that involves using a fitted model to impute a single value of missing values based on observed ones. RI could offer a pragmatic alternative in the context of prediction, that is simpler to apply at the point of prediction. Moreover, the use of missing indicators has been proposed to handle informative missingness, but it is currently unknown how well this method performs in the context of CPMs.

## 1.2 Methods

We performed an extensive simulation study where data were generated under various missing data mechanisms to compare the predictive performance of CPMs developed using both imputation methods. We consider deployment scenarios where missing data is permitted or prohibited, and develop models that use or omit the outcome during imputation and include or omit missing indicators. We assume that the missingness mechanism remains constant between the stages of the model pipeline.

## 1.3 Results

When complete data must be available at deployment, our findings were generally in line with widely used recommendations; that the outcome should be used to impute development data when using MI, yet omitted if using RI. When imputation is applied at deployment, developing a model that instead omits the outcome from imputation at development was preferred. Missing indicators were found to improve model performance in some specific cases, but found to be harmful when missingness is dependent on the outcome.

## 1.4 Conclusion

We provide evidence that commonly taught principles of handling missing data via MI may not apply to CPMs, particularly when data can be missing at deployment. In such settings, RI and missing indicator methods can (marginally) outperform MI. As shown, the performance of the missing data handling method must be evaluated on a study-by-study basis, and should be based on whether missing data are allowed at deployment. Some guidance is provided.

## 2 Background

Clinical Prediction Models (CPMs) can be used to guide clinical decision making and facilitate conversations about risk between care providers and patients[1]. A CPM is a mathematical tool that takes patient and clinical information (predictors) as inputs and, most often, produces an estimated risk that a patient currently has (diagnostic model) or will develop (prognostic model) a condition of interest[2]. A common challenge in the development, validation and deployment of CPMs is the handling of missing data on predictors and outcome data. The methods used to handle missing data in CPM development and validation are usually complete case analysis or multiple imputation (MI) approaches[1,3], the latter of which is often heralded as the gold standard in handling missing data. This topic has recently received renewed interest, with authors arguing the basis on which MI is used relies too heavily on principles relevant to causal inference and descriptive research, which are less relevant when the goal is to provide accurate predictions[4].

The objectives of prediction research differ from those of descriptive or causal inference studies. For the latter, missing data should be handled in such a way that minimises bias in the estimation of key parameters, and generally this is achieved through multiple imputation of missing data. In the development of prediction models, however, unbiased parameter estimates are not necessarily the ones that optimise predictive performance. Moreover, in prediction research we must distinguish between handling missing data across the entire model pipeline; model development, model validation, and model deployment (or prediction time), and anticipate whether missing data shall be allowed at deployment. Ideally, all predictors considered for inclusion in a CPM should be either readily available, or easily measured, at the point of prediction. There exist, however, notable examples that allow missingness at the point of prediction[3]; the QRisk3[5] and QKidney[6] algorithms are examples of such models that allow users to make a prediction in the absence of clinical predictors (such as cholesterol) that may not be available, or easily measured, at the time of prediction.

Best practice states that the outcome should be used in the imputation model when applying MI[7], creating a congenial imputation model. Clearly the outcome is unknown at prediction time, and applying imputation without the outcome would violate the assumption of congeniality. Since model validation should evaluate predictive performance under the same missing data handling strategy to be used in practice, the outcome should be omitted from any imputation model at validation, potentially resulting in less accurate imputations since predictors are normally predictive of the outcome. We therefore define "performance under no missingness", where we assume all predictors are always available (or easily obtained) at prediction time, and "performance under missingness", assuming missing data is allowed and will be imputed at deployment.

Regression imputation (RI) could provide a more pragmatic alternative to MI in the context of prediction. RI is a form of single imputation that fits a model to impute missing predictors using observed data. The key difference between RI and MI is that RI is based on a single equation, and produces one imputed value (deterministic process), whereas MI is a stochastic sampling process that involves repeatedly sampling from a distribution. For RI to be applied in practice, only the imputation model(s) needs to be available alongside the full prediction model, as opposed to MI which generally also requires access to the development dataset. Existing literature has, however, demonstrated several pitfalls of RI in the context of causal estimation - it is highly sensitive to model misspecification, can increase correlation between predictors and underestimate variability in parameter estimates[8]. Although these issues may therefore persist within the prediction context, they may not apply since the recovery of unbiased parameter estimates is no longer of direct concern. RI may also overcome some of the previously mentioned issues related to predictive modelling with MI, since inclusion of the outcome in the imputation model is not recommended[1]. To our knowledge these issues and challenges have not been studied to date.

Both MI and RI are techniques devised under the assumption that data are missing at random, i.e. missingness does not depend on unobserved values. The validity of the MAR assumption within health data is often dubious, especially when using routinely collected data[9,10], however these definitions were created with the goal of recovering unbiased parameter estimates in mind and therefore may be less relevant to the prediction modelling context[4]. Within routinely collected data, the recording of key clinical markers is often driven by the needs of the patient and clinical judgments of the care provider[10]. Missingness is therefore potentially informative with respect to a patient's current or future condition, and including information about the way an individual has been observed into a prediction model has the potential to improve its predictive performance[11]. A commonly used, effective approach to achieve this is through the inclusion of missing indicators as predictors in a CPM.

This study therefore aims to explore the use of missing indicators as model predictors alongside both regression and multiple imputation. We explore the effect of omitting/including the outcome from each imputation model at development, and imputing data without the outcome at validation (and therefore deployment). We compare the two imputation strategies under each development/validation strategy. Our results will inform recommendations on the handling of missing data during model development and deployment that will be especially relevant to applied researchers developing clinical prediction models.

# 3    Methods

We performed an extensive simulation study in which we evaluated a range of different missingness mechanisms. Our study has been designed according to best practice and reported according to the ADEMP structure (modified as appropriate for a prediction-focused study), proposed by[12].

## 3.1    Aims

To compare MI and RI approaches in imputing missing data when the primary goal is in developing and deploying a prediction model, under a range of missing data mechanisms (MCAR, MAR, MNAR), with/without a missing indicator and with/without the outcome included in the imputation model. Each of these will be examined both allowing for and prohibiting missing data at deployment, and performance will be estimated separately for each of these two scenarios.

Throughout this study, we assume that both the missingness mechanism and handling strategy will remain the same across validation and deployment, and therefore validation is a valid replication of model deployment and our performance estimates are reliable estimates of model performance at deployment. The only case where this is not true is when we impute data using the outcome at validation, which will be discussed in more detail in the following sections.

## 3.2 Data-generating mechanisms

We focus on a logistic regression-based CPM to predict a binary outcome, $Y$, that is assumed to be observed for all individuals (i.e. no missingness in the outcome) during development and validation of the model. Without loss of generality, we assume that the data-generating model contains up to three predictors, $X_1$, $X_2$ and $U$, where $X_1$ is partially observed and potentially informatively missing (depending on simulation scenario, as outlined below), $X_2$ is fully observed and $U$ is unobserved. We denote missingness in $X_1$ with binary indicator $R_1$, where $R_1 = 1$ if $X_1$ is missing, and $R_1 = 0$ if it is observed.

We construct four separate DAGs depicted in Figure 3.1, each representing different missingness structures covering: Missing Completely at Random (MCAR), Missing at Random (MAR), Missing Not at Random dependent on $X_1$ (MNAR-X) and Missing Not At Random dependent on Y (MNAR-Y). MNAR-Y is often instead referred to as "MAR-Y", however we felt that in this instance, classifying this mechanism as missing not at random is more appropriate since the outcome ($Y$) is not observed at the point of prediction. This mechanism could therefore be classified differently according to which stage of the CPM pipeline we are referring to: at model development (and equally at model validation) it may be more appropriate to refer to this as "MAR-Y" since information on observed outcomes is available at this stage. We adopt a single name across this study, however, to ease presentation and interpretation of results.

The DAGs further illustrate how missingness in $X_1$ is related to $X_1$ or $X_2$. In each of the DAGs, $X_1^*$ represents the observed part of $X_1$ and $R_1$ is the missing indicator.

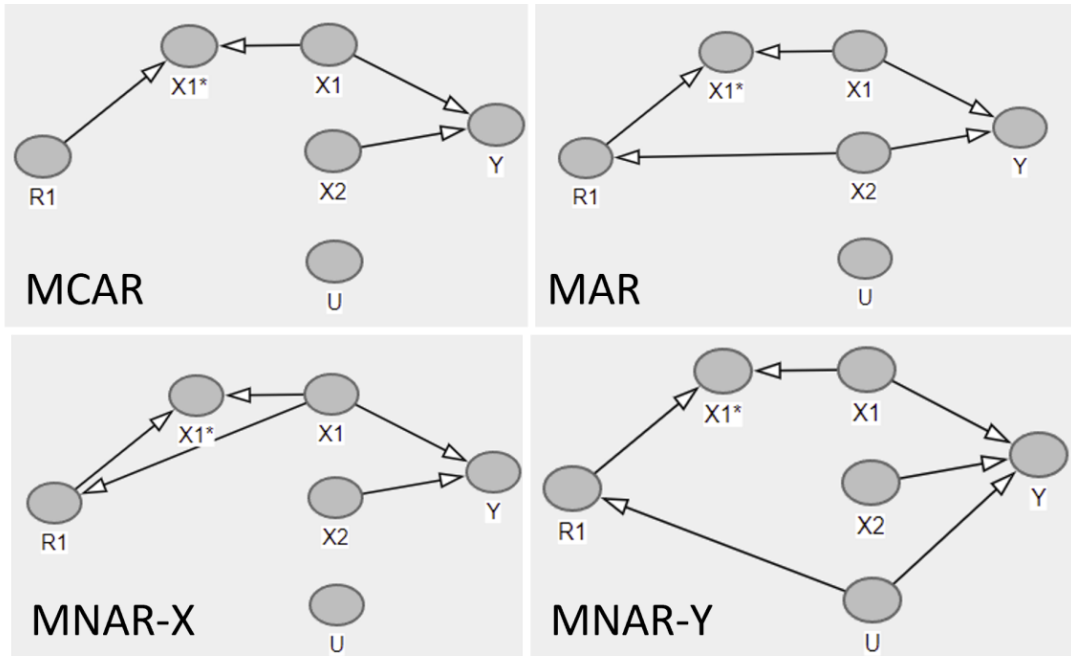

*Figure 3.1: Directed Acyclic Graphs for four missingness structures, constructed via our data generating mechanisms*

In order to reconstruct these DAGs in simulated data, we stipulate the following parameter configurations:

$X_1$ and $X_2$ are drawn from a bivariate normal distribution to allow moderate correlation between the two predictors, such that:

$$\mathbf{X} \sim MVN(\boldsymbol{\mu}, \Sigma)$$

Where $\boldsymbol{\mu} = \begin{bmatrix} 0 \\ 0 \end{bmatrix}$

and $\Sigma = \begin{bmatrix} 1 & 0.4 \\ 0.4 & 1 \end{bmatrix}$

- $R_1 \in \{0,1\}$, and $P[R_{1i} = 1] = expit(\beta_0 + \beta_{X_1}X_{1i} + \beta_{X_2}X_{2i} + \beta_Y Y_i)$ i.e. missingness in $X_1$ can depend on $X_1$, and/or $X_2$, and/or $Y$.

- $\beta_{X_1}, \beta_{X_2}$ and $\beta_Y$ were varied across $\{0, 0.5, 1\}$

- $Y$ is binary, with $P[Y_i = 1] = expit(\gamma_0 + \gamma_{X_1}X_{1i} + \gamma_{X_2}X_{2i} + \gamma_{X_1 X_2}X_{1i}X_{2i})$

- $\gamma_{X_1}$ and $\gamma_{X_2}$ were varied across $\{0.5, 0.7\}$

- $P[Y = 1] = \pi_Y$ was fixed to be 0.1.

- $P[R_1 = 1] = \pi_{R1} \in \{0.1, 0.25, 0.5, 0.75\}$

- $\gamma_{X_1 X_2}$ can take values $\{0, 0.1\}$.

- $\beta_0$ and $\gamma_0$ are calculated empirically as required to set the desired level of $\pi_{R1}$ and $\pi_Y$.
  $U$ was not directly simulated, but the MNAR-Y scenario generated via the inclusion of $\beta_Y \neq 0$.

The parameter values above were selected to represent what might be observed in real-world data, assuming that $X_1$ and $X_2$ can represent some summary of a set of model predictors. The $\gamma$ coefficients (specifying the relationship between predictors and the model outcome) are strong, as we wish to illustrate the impact of (potentially informative) missingness in a very important predictor (or set of predictors).

Datasets will be generated with $n = 10000$ records from the DGMs described above, and split 50/50 into development and validation sets. The development and validation sets will therefore contain 5000 records each. This is chosen as a suitably large size that should be sufficient to estimate underlying parameters, and avoid overfitting. It is also in line with what we would expect to see in electronic health record data.

Each simulated DGM will be repeated for 200 iterations. The parameter values listed above result in a total of 864 parameter configurations. 200 iterations was selected as optimal to balance the requirement to obtain reliable estimates of key performance metrics (by using a sufficient number of repetitions) with the size of the study and computational requirements of repeatedly running multiple imputation over a large number of simulated scenarios.

Our simulation procedure first generates data under the DGMs described Figure 3.1, according to the above parameter configurations. We then take a split-sample approach to assessing model performance (we recognise that this is a statistically inefficient approach to use in model development applications, but our simulated sample size is sufficiently large that this should not pose an issue)[13]. We randomly separate the data into 50% development and 50% validation, fit the models on the development data, and calculate performance measures on the derived models applied to the validation set. The full simulation procedure is illustrated in Figure 3.2. Note that in this instance, we fit the imputation models separately in the development and validation sets. Since the DGMs and missingness mechanisms remain constant between the two datasets, we assume that the fitted imputation model would not change and this is therefore a valid approach to take. In a real-world setting, however this would not be feasible as only a single patient's data would be available at the point of prediction, and we would want to use the same imputation model as was used/developed in the development data.

## 3.3  Missing data handling strategies

We consider two main methods for handling missing data at the development and implementation stages of the CPM pipeline: multiple imputation and regression imputation. Multiple imputation can be applied with relative ease at the model development stage, specifying an imputation model and method for every predictor with missing data (in this case just $X_1$), conditional on other data available at model development/implementation. Multiple draws are then made from the imputation model, resulting in multiple completed datasets. The relevant CPMs are then fit separately to each resulting imputed dataset, and the model coefficients pooled according to Rubin's rules to obtain a single set of model coefficients. For regression imputation, we follow a similar process in fitting a model to the missing predictor(s) based on observed data. The key difference between MI and RI is that we obtain only a single completed dataset under RI. We can then fit the analysis model to this new complete data to obtain the CPM's parameter estimates. Both methods can be implemented using the `mice` package in R (amongst others), but more flexible user-defined imputation models for regression imputation could be fit with relative ease using alternative modelling packages. In this study, we consider non-stochastic RI, as we only have a single missing predictor so there is no requirement to apply sampling procedures to handle more complex patterns of missingness in the model predictors.

Applying multiple imputation to incomplete data for new individuals at the point of prediction is more challenging as it is not generally possible to extract the final imputation model from the output provided by standard statistical software. In order to "fix" the imputation model for new individuals, it has therefore instead been proposed that the new individual's data should first be appended to the original (imputed) development data, and the imputation re-run on the new stacked dataset[14–16]. Regression imputation, on the other hand, is easier to implement at the point of prediction, since flexible models can be defined for each (potentially) missing predictor, and these models can be stored and used to impute at the point of prediction for new individuals. Ideally, model validation should follow the same steps as model deployment in order to properly quantify how the model will perform in practice. However, since validation is usually completed for a large cohort of individuals at the same time (as opposed to a single individual), it is likely that missing data imputation would take place as a completely separate exercise, with the imputation model depending solely on the validation data.

In this study, we take a split-sample approach to model validation, whereby the imputation is run (and therefore separate imputation models fit) within the development and validation data. This would not be possible in practice in order to predict for new individuals, since only data for a single individual would be available at the point of prediction. We expect, however, that the fitted imputation model(s) would remain the same between the development and validation datasets under this simulation framework, since the DGM and missing data mechanisms remain constant between the two datasets.

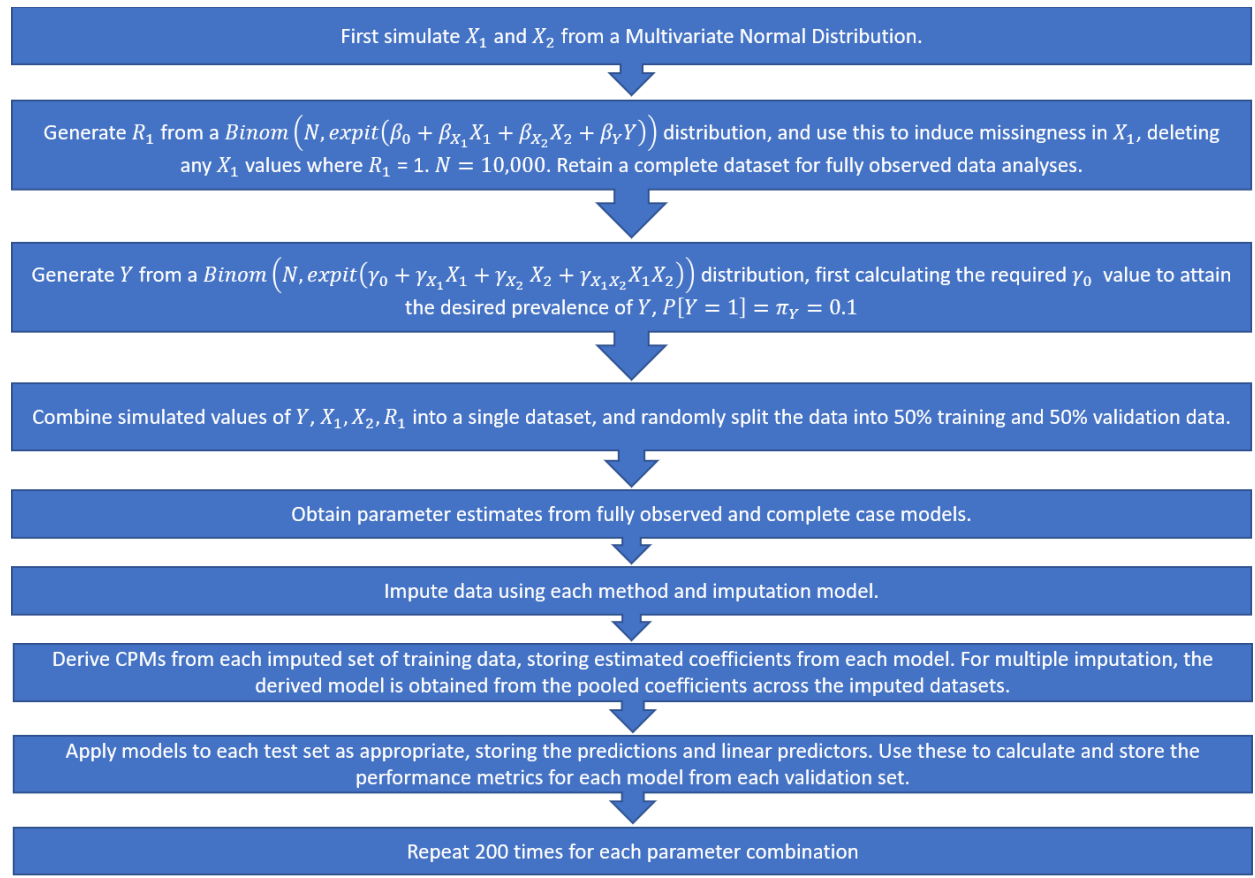

Figure 3.2: Simulation procedure, step-by-step

## 3.4 Fitted CPMs

We fit three possible CPMs to the development data (under each different imputation method), firstly with a simple model including both derived predictors and their interaction. We then fit models incorporating missing indicators, as well as considering a model with an interaction between the missing covariate $X_1$ and its missing indicator $R_1$[17]:

- Predictors and their interaction only: $P[Y_i = 1] = expit(\gamma_0 + \gamma_{X_1}X_{1i} + \gamma_{X_2}X_{2i} + \gamma_{X_1X_2}X_{1i}X_{2i})$

- Inclusion of an additional missing indicator: $P[Y_i = 1] = expit(\gamma_0 + \gamma_{X_1}X_{1i} + \gamma_{X_2}X_{2i} + \gamma_{X_1X_2}X_{1i}X_{2i} + \gamma_{R_1}R_{1i})$

- Inclusion of an interaction between the $X_1$ and $R_1$ terms in the outcome model:
  $P[Y_i = 1] = expit(\gamma_0 + \gamma_{X_1}X_{1i} + \gamma_{X_2}X_{2i} + \gamma_{X_1X_2}X_{1i}X_{2i} + \gamma_{R_1}R_{1i} + \gamma_{R_1X_1}R_{1i}X_{1i})$

Each model will be derived on incomplete data using both MI and RI, with and without the outcome in the imputation model. The derived models will then be applied to the validation set according to the strategies listed in Table 3.1.

The underlying imputation model (for missing predictor $X_1$) will be the same for both regression imputation and multiple imputation, as they are both based on a linear regression model. The key difference between the two methods (for the implementation used in this study) is that multiple imputation allows for multiple draws from this same imputation model, allowing the parameter estimates to account for the uncertainty associated with the missing data handling strategy.

The "passive imputation" approach is used here to account for the $X_1:X_2$ interaction in the analysis model, whereby a value ($X_1^*$) is imputed for $X_1$, and $X_1:X_2$ is calculated as $X_1^*:X_2$. This results in an imputation model that is not congenial with the analysis model which can result in biased parameter estimates. Due to the lack of availability of the outcome $Y$ at the point of prediction, any imputation model (based on MI) will also be uncongenial at the point of prediction, therefore achieving congeniality was not a primary consideration in the implementation of MI in this particular case. Other methods could, however, have been considered to mitigate any bias introduced by passively imputing the $X_1:X_2$ interaction term such as the "just another variable" (JAV) approach, which separately imputes a value for $X_1:X_2$ as part of the imputation procedure. The JAV approach has been shown to perform slightly better than passive imputation under some circumstances.

Table 3.1: Table 1: Imputation and validation strategies

| Strategy | Description | Missingness Allowed at Deployment |
| --- | --- | --- |
| $DA + VA$ | All data, before inducing missingness, at development and validation | No |
| $DY + VY$ | With Y in imputation model at development and validation | No |
| $D\bar{Y} + V\bar{Y}$ | Without Y in imputation model at development and validation | Yes |
| $DY + V\bar{Y}$ | With Y in imputation model at development, but not at validation | Yes |
| $DY + VA$ | With Y in imputation model at development, all (completed) data required at validation | No |

| $D\bar{Y} + VA$ | Without Y in imputation model at development, all (completed) data requied at validation | No |

## 3.5 Development/Validation scenarios

We apply each of the imputation and validation strategies described above in Table 3.1, $DA + VA$ to $D\bar{Y} + VA$. MI is performed using Bayesian linear regression as the underlying form of the imputation model (as implemented by mice.impute.norm in the mice package in R) and 20 imputed datasets. Parameter estimates are pooled across the the imputed datasets according to Rubin's rules, and, similarly, we take the "pooled performance" approach to validation[18] whereby imputation-specific predictions are obtained in the multiply imputed validation datasets, the predictive performance of each imputed dataset is calculated, and then these estimates of model performance are pooled across the imputed datasets. An alternative strategy would be to first pool the predictions obtained from the validation set, obtaining a single measure of predictive performance for each validation dataset (so-called "pooled predictions"). The chosen "pooled performance" strategy was based on recommendations from[18], as "pooled predictions" can at times over-estimate model performance. It is, however, perhaps unrealistic to expect that multiple predictions would be obtained when applying a model in practice and therefore the "pooled predictions" method may make more sense when trying to validate the model as it will be used in a real-world setting.

$DY + VA$ and $D\bar{Y} + VA$ can be considered estimates of "performance under no missingness", and to estimate this we retain the fully observed validation data before missing data are induced. $DY + VY$ will be classed as "approximated performance under no missingness", since it attempts to estimate performance assuming no missingness at deployment, but with missing data in the validation set. Note, however, that this strategy could not realistically be applied in a real-world setting (at prediction/implementation time) due to the inclusion of $Y$ during imputation of the validation data.

$D\bar{Y} + V\bar{Y}$ and $DY + V\bar{Y}$ are both strategies that could be applied in practice when missingness is allowed, with the key difference between the two being that Y is omitted from the imputation model at validation and deployment. They therefore correspond to measures of "performance under missingness", assuming this imputation strategy could reasonably be applied at the point of prediction. For this approach, we do not have a true estimand that we are targeting, so the methods will be compared against each other to establish the optimal missing data handling strategy.

The fully observed data strategy in $DA + VA$ will be considered to be the reference approach, since this is equivalent to the data-generating model prior to applying any missing data strategy, and will be used as a comparator for other methods. Strategies that do not allow missingness at deployment will be directly compared against this approach, as will $DY + VY$ since it aims to approximate performance under no missingness.

In strategies $DA + VA$, $DY + VY$, $D\bar{Y} + V\bar{Y}$, we assume that the missingness mechanism, the missing data strategy, and proportion of missing data remain constant between model development and validation, which is perhaps a strong assumption in practice. For strategies that allow missingness at deployment, we assume that the missingness mechanism and proportion of missingness remain constant across the pipeline.

## 3.6 Target and Performance Measures

Our key target is an individual's predicted risk, and we compare each method's ability to estimate this using the following metrics of predictive performance, covering both calibration and discrimination[1,19]:

- Calibration-in-the-large (CITL) - the intercept from a logistic regression model fitted to the observed outcome with the linear predictor as an offset

- Calibration slope - the model coefficient of the linear predictor from a model fitted to the observed outcome with the linear predictor as the only explanatory variable

- Discrimination (Concordance/C-statistic) - a measure of discriminative ability of the fitted model. Defined as the probability that a randomly selected individual who experienced the outcome has a higher predicted probability than a patient that did not experience the outcome

- Brier score - a measure of overall predictive accuracy, equivalent to the mean squared error of predicted probabilities

We assume that the estimates of the above measures are valid representations of performance at model deployment, based on the following assumptions: 1) when missingness is allowed at deployment, it will be imputed in the same way as performed in our validation set, 2) the missingness mechanism will not change between validation and deployment, and 3) when missingness is not allowed at deployment, we assume that the data-generating mechanism remains constant across validation and deployment.

We also extract the obtained parameter estimates and any associated bias from each fitted CPM, as these will likely provide insight into the performance of the models.

## 3.7 Software

All analyses are performed using R version 3.6.0 or greater[20]. The pROC library[21] was used to calculate C-statistics and the mice package[22] was used for all imputations. Code to replicate the simulation can be found in the following GitHub repository:

https://github.com/rosesisk/regrImpSim.

## 4 Results

Select parameter combinations have been chosen to highlight important results within this chapter, but full results for all combinations are made available in a rShiny dashboard at https://rosesisk.shinyapps.io/regrimpsim.

## 4.1 Predictive Performance

Dashed lines mark the median performance metric in the complete data for the Brier Score (Figure 3) - i.e. before any missingness is induced. For the Calibration slope (Figure 4), vertical lines are placed at 1.

For simplicity, we restrict our results to a single parameter configuration for each missingness mechanism. The following parameters remain fixed throughout this section: $\gamma_{X1} = \gamma_{X2} = 0.7, \gamma_{X1X2} = 0.1, \pi_{R1} = 0.5$.

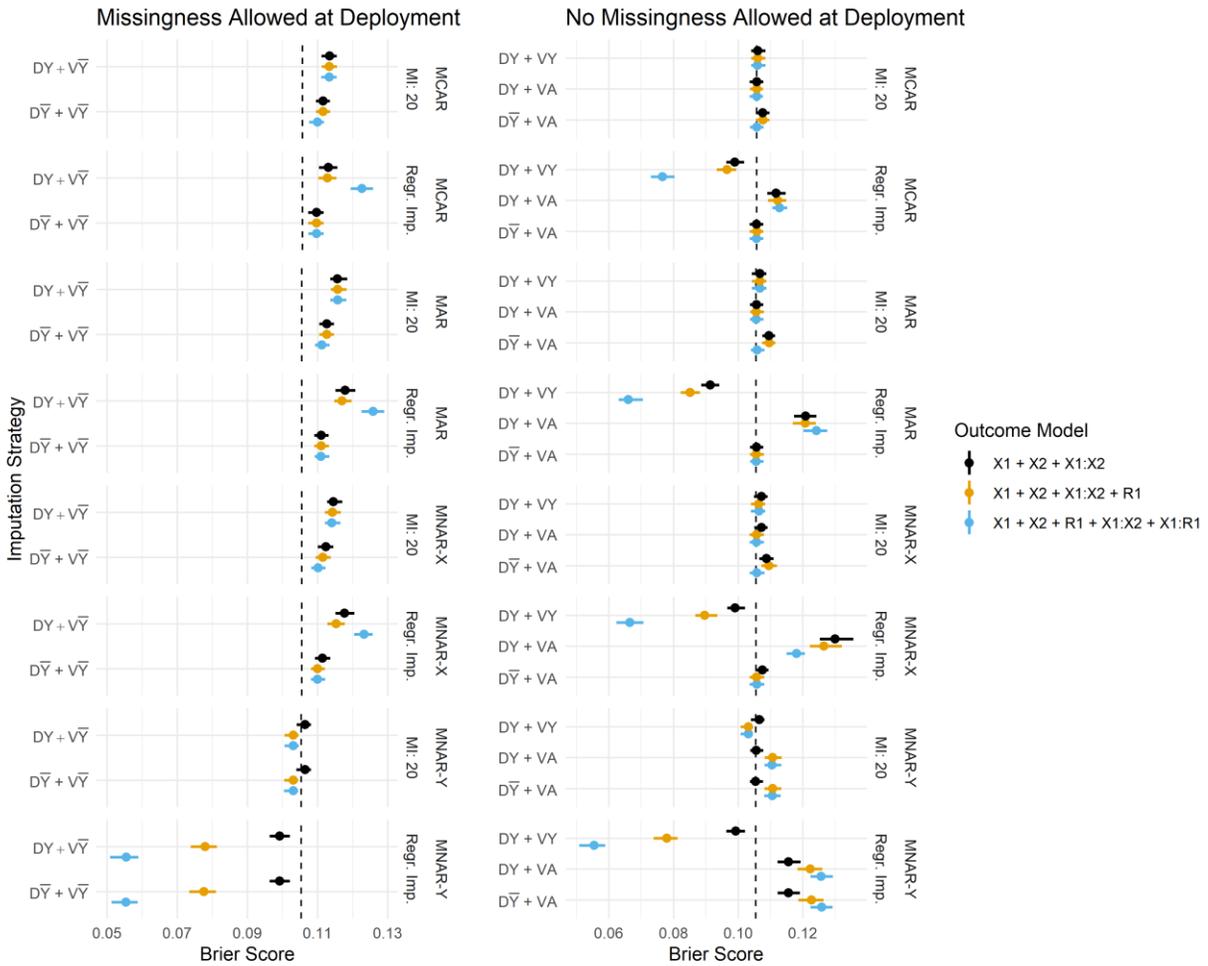

Figure 4.1: Brier Score estimates across Development/Validation scenarios, imputation methods and missingess mechanisms. The vertical dashed lines represent estimates from the complete data scenario (DA + VA)

### 4.1.1 Inclusion of the Outcome in the Imputation Model

Figure 4.1 summarises the estimated Brier Scores for each imputation strategy (defined in Table 3.1) for both imputation methods and all fitted outcome models, and calibration slopes are presented in Figure 4.2 The imputation strategies have been split according to whether or not they allow missingness at model deployment.

When missingness is allowed at deployment (i.e. imputation will be applied at the point of prediction), we primarily want to know whether imputation should be performed with or without the outcome at development, since at deployment it must be omitted from the imputation model. We observe that predictions in the validation set are far better calibrated under $D\bar{Y} + V\bar{Y}$ than $DY + V\bar{Y}$, i.e. when the imputation model remains consistent between development and validation/deployment (Figure 4.2). This difference can also be seen in marginal differences in the Brier Score (Figure 4.1).

When complete data is required at deployment, RI still performs better under $D\bar{Y} + VA$ than $DY + VA$ in terms of calibration (Figure 4.2), and this result becomes more pronounced when data are MAR or MNAR-X at development. MI, on the other hand, favours retention of $Y$ in the imputation model at development in terms of both Brier Score and model calibration (under no missingness at deployment). Figure 4.3 summarises the parameter estimates from the fitted models, and we can see that effect estimates are less biased for RI: No Y and MI: Y, which is in line with the predictive performance estimates.

A notable result is that the $DY + VY$ strategy fails to recover the performance under no missingness ($DY + VA$) under both methods, i.e. when $Y$ is used to impute at both development and validation. Under MI, this strategy should be a valid means of estimating the performance under complete data at deployment, but there are marginal differences in the calibration slope between MI: $DY + VY$ and MI: $DY + VA$. Moreover, the performance of RI considerably breaks down with the inclusion of $Y$ in the imputation model under all missingness mechanisms and regardless of whether missingness is allowed at deployment. This same result is evident in Figure 4.3): Parameter estimation, where RI: Y consistently fails to recover unbiased parameter estimates.

### 4.1.2 Comparison of Imputation Methods

Overall, the performance estimates from MI are considerably more stable than those from RI; the differences in Brier Score between the various imputation strategies are smaller for MI as can be seen from Figure 4.1. When RI performs poorly, the poorer model tends to be even worse than the worst MI model, in terms of both Brier Score and calibration. RI does, however, often perform at least as well as, or better than, MI when the preferred imputation model is applied (i.e. omitting the outcome when applying RI). For example, both methods perform comparably under $D\bar{Y} + V\bar{Y}$. With missingness permitted at deployment, performance is comparable between MI: $DY + VA$ and RI: $D\bar{Y} + VA$.

### 4.1.3 Inclusion of a Missing Indicator

The inclusion of a missing indicator appears to have minimal impact on the Brier Score and calibration under most methods and imputation strategies, with a few notable exceptions.

#### 4.1.3.1 Missingness allowed at deployment

Under MNAR-Y and missingness allowed at deployment, inclusion of a missing indicator in the outcome model provides reductions in the Brier Score, and improvements in the C-statistic (for both imputation methods, C-statistics presented in the supplementary materials, performing even better than the complete data model, since the inclusion of the indicator allows the CPM to extract additional information about the outcome that is not available from only observed data. We can further see from Figure 4.3 (parameter estimates) how the estimates of the intercept ($\hat{\gamma}_0$) are less biased for both methods when the indicator is included under this mechanism.

Under this same missingness mechanism (MNAR-Y), inclusion of the indicator and its interaction with $X1$ produce marginally overfit models for RI (Calibration slope < 1, Figure 4.2). This result is explored further in the supplementary materials, where we present plots of the predicted risk distributions - we see that this method produces predictions that are very close to 0 and 1.

### 4.1.3.2 No missingness allowed at deployment

Inclusion of the indicator corrects the CITL for both MI: $DY + VA$ and RI: $D\bar{Y} + VA$ under MAR and MNAR-X structures when missingness is not allowed at deployment (in Supplementary material). Further improvements in the calibration slope can be achieved through inclusion of the additional $X_1 : R_1$ term under the preferred imputation model for both methods. Conversely, under MNAR-Y and fully observed data at deployment, inclusion of a missing indicator results in underestimated average predicted risk (CITL > 0) for both imputation methods, and severe overfitting for RI (calibration slope < 1, Figure 4.2). Models developed using MI under MNAR-Y, are, however generally still well calibrated (slope close to 1) whether or not missing data are allowed at deployment.

Interestingly, inclusion of the $X_1:R_1$ interaction term where only complete data will be used at deployment appears to (marginally) improve the calibration slope for MI under all missingness mechanisms. Clearly, this term would be 0 for all new individuals however its inclusion at development seems to aid model performance at deployment.

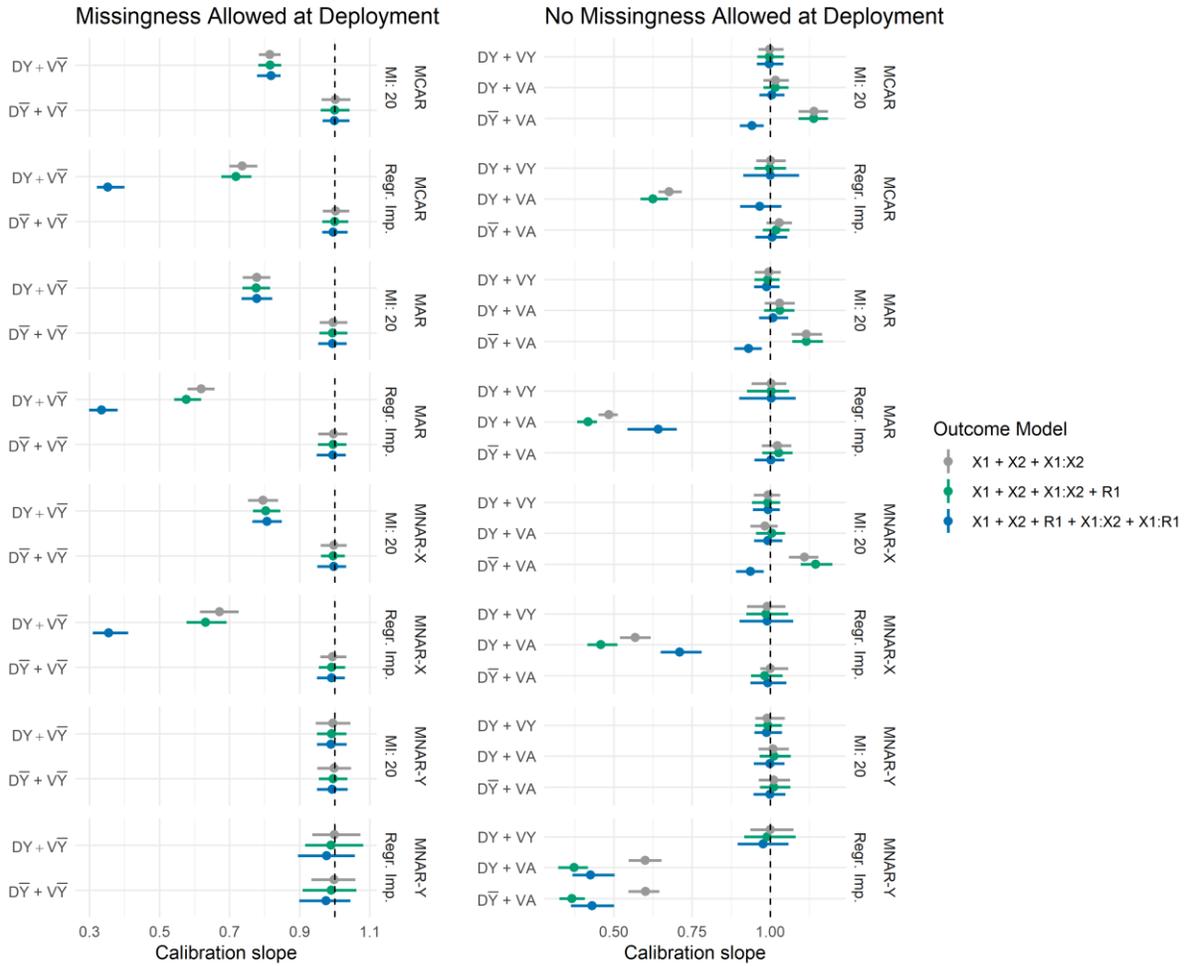

*Figure 4.2: Calibration Slope estimates across imputation strategies, imputation methods and missingness mechanisms*

## 4.2 Parameter Estimation

We further present results for the CPM parameter estimates for selected scenarios presented in Figure 1. The same parameter configurations as specified in previous sections are used here.

Presented are the coefficients obtained from fitting each model within the development data. For RI, we present coefficients pooled according to Rubin's rules.

RI using the outcome in the imputation model consistently produces parameter estimates that are both biased and much larger in magnitude than any other method, and this is reflected in the predictive performance estimates - this method fails to produce models that are well calibrated or have good overall accuracy (Brier Score). We frequently observe Brier Scores (Figure 4.1) that are too extreme under $DY + VY$, and at times calibration slopes that suggest the predicted risks are too extreme. Both of these are indicative of overfitting and likely due to the relative size of the parameter estimates compared to other methods. RI: $D\bar{Y}$ generally estimates parameters with minimal bias, with the exception of under MNAR-Y where, similarly, parameter estimates are large and generally very biased. This is further reflected in the Brier Scores (Figure 4.1) and predicted risk distribution in supplementary material).

Perhaps as expected, MI: $D\bar{Y}$ fails to recover unbiased parameter estimates under all missingness mechanisms, whereas MI: Y generally recovers the true parameter values well. It can therefore be observed that biased parameters do not necessarily result in worse predictive performance, as models developed under MI: No Y, and missing data imputed in this same way at validation/deployment were favoured over models developed using the outcome at development.

The inclusion of a missing indicator reduces bias in the estimate of the effect of $X1$ on $Y$ ($\gamma_{X1}$) under both MNAR structures for MI: Y.

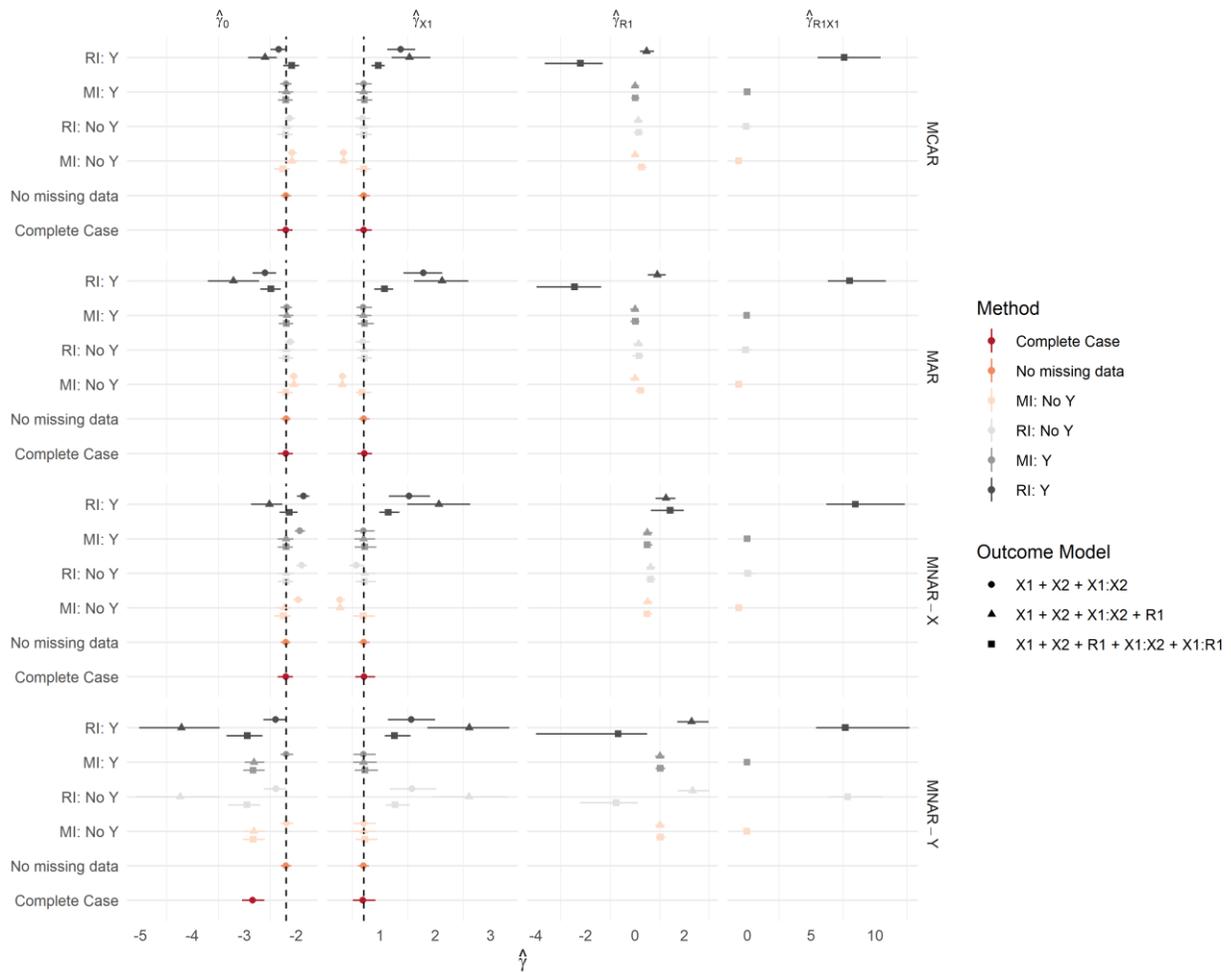

*Figure 4.3: Parameter estimates across all missingness mechanisms. Missingness fixed at 50%*

# 5    Discussion

In this study, we have assessed model performance for multiple imputation and regression imputation, with and without the use of missing indicators across a range of missingness mechanisms. We considered how/when the outcome should be used in the imputation model for missing covariates, whether RI could offer a more practical and easier to implement solution than MI, and how the inclusion of missing indicators affects model predictive performance. All of these questions were considered in relation to whether or not missing data will be allowed once the model is deployed in practice. We have provided a concise list of recommendations in Table 5.1.

In the context of recovering unbiased parameter estimates, the literature advocates the use of the observed outcome in the imputation model for MI[7]. In the context of predictive performance, we found that RI consistently performed better when $Y$ is instead omitted from the imputation. This strategy is recommended by[23] for RI, where the author notes that including the outcome in the imputation model artificially strengthens the relationship between the predictors and outcome. MI overcomes this issue by introducing an element of randomness to the imputation procedure.

We further observed that the performance of a model with inconsistent imputation models between development and validation ($DY + V\bar{Y}$) performed worse than one where the imputation model remained consistent. For instance, omitting the observed outcome at both stages resulted in better predictive performance, even when using MI. Although we have also observed that the inclusion of $Y$ in the imputation model helps in recovering unbiased effect estimates, others have recommended a more considered approach when targeting a model that allows for missing data at prediction time. In a simulation study conducted by[15], multiple imputation including the outcome produced larger overall prediction error than omitting the outcome entirely, as the out-of-sample imputations were biased by attempting to use the imputation model derived during development (with $Y$) to impute in the test set (where $Y$ is unobserved). This imputation bias carried through to the overall performance of the model. This is in line with our findings, whereby a consistent imputation model between development and deployment resulted in stronger performance overall, at the cost of slightly biased parameter estimates. An interesting result to note here is that a model with unbiased model parameters is not necessarily the one that predicts best, especially when data will be imputed again at deployment.

We have demonstrated that RI could offer a practical alternative to multiple imputation within the context of prediction. As discussed above, there are several challenges associated with applying multiple imputation during deployment of a CPM, including but not limited to: requiring access to the development data and the availability of computational power and time. Recent developments, have, however proposed methods that potentially mitigate these requirements[16]. RI also overcomes both of these major issues, in that only the deterministic imputation models would be required to produce imputations during model implementation. We emphasize, however, that RI consistently showed poor performance when the observed outcome was included in the imputation model, and this method should therefore only be used when other model covariates are used to impute missing ones. MI, on the other hand, proved to be more stable across a range of scenarios and imputation models.

The careful use of missing indicators has also proven to be beneficial in specific cases. For example, under MNAR-X, multiple imputation has marginally stronger performance in both imputed and complete deployment data when a missing indicator is included in the outcome model. Under incomplete data at deployment, inclusion of an indicator further provided small improvements in overall predictive accuracy to MI under MNAR-Y, but resulted in some overfitting for models developed and applied using RI. Since MI is only assumed to recover unbiased effects under MAR, the indicator appears to correct this bias under informative missingness patterns. We observe further (marginal) improvements in model performance for MI with the inclusion of an interaction between $X_1$ and $R_1$ when missingness is not allowed at deployment, and when Y is omitted from the imputation model at development ($D\bar{Y} + VA$). However, we noted some surprising results in the use of indicators when data are MNAR-Y; specifically when missing data are not allowed at deployment the inclusion of the indicator is harmful and resulted in small increases in the Brier Score, and poor Calibration-in-the-large.

Table 5.1: Table of recommendations for the use of multiple imputation, regression imputation and missing indicators in the development and deployment of clinical prediction models

| **Recommendations** |
| --- |
| Assuming no missing data will be present at deployment, multiple imputation (including the outcome) is recommended as the best strategy |
| Where missingness is allowed at deployment, and multiple imputation is impossible at the point of prediction, regression imputation can be used as an alternative. |
| Always omit the outcome from the imputation model under regression imputation. |
| Where data are assumed to be MNAR-X or MNAR-Y, inclusion of a missing indicator can offer marginal improvements in model performance, and does not harm performance under MCAR or MAR mechanisms |
| The use of missing indicators under MNAR-Y can harm model performance when missingness is not allowed at deployment, and is not recommended |

Related literature under a causal inference framework by Sperrin et al[4] and Groenwold et al[24] has found that the inclusion of missing indicators is not recommended under MCAR, and can lead to biased parameter estimates under this missingness structure. Van Smeden et al[25] discuss at length how missing indicators should be approached with caution in predictive modelling - inclusion of a missing indicator introduces an additional assumption that the missingness mechanism remains stable across the CPM pipeline; an assumption that is generally dubious, but especially within routinely collected health data. The propensity to measure certain predictors is likely to vary across care providers, settings and over time as clinical guidelines and practices change. This in turn potentially changes the relationship between the missing indicator and the outcome and could have implications for model performance. As others have highlighted, the strategy to handle missing data should be devised on an individual study basis, taking into consideration the potential drivers of missingness, how stable these are likely to be, and how/whether missing data will be handled once the model has been deployed.

We recommend that the strategy for handling missing data during model validation should mimic that to be used once the model is deployed, and that measures of predictive performance be computed in either complete or imputed data depending on whether missingness will be allowed in the anticipated deployment setting or not. For example, complex model applications integrated into electronic health record systems are better suited to applying imputation strategies, whereas simple models that must be filled in at the point-of-care are more likely to require a complete set of predictors. The difference between performance allowing for and prohibiting missing data at deployment may also be of interest in assessing any drop in performance related to the handling of missing data. Interestingly, we have observed somewhat different results depending on whether missingness is allowed at deployment or not. We may therefore wish to optimize a model for either one of these use cases, resulting in a different model (and hence different coefficients) dependent on whether we envisage complete data at implementation.

Although we have considered a wide range of simulated scenarios, a key limitation to this study is that we only considered a relatively simple CPM with two covariates, where only one was allowed to be missing. This was to restrict the complexity and size of the work, as only a limited set of scenarios can realistically be presented. We do, however, expect that the fundamental findings would generalise to more complex models, since we could consider each of the two model predictors to represent some summary of multiple missing and observed predictors. With more predictors in the model, there would not be any additional missingness mechanisms and we therefore anticipate that such complex models would not provide any additional insight. A further possible limitation is that this work has been restricted to the study of a single binary outcome, although we would not expect the results to change in the context of e.g. continuous or time-to-event outcomes. We accompany this work with a rShiny dashboard allowing readers to explore our results in more detail across the entire range of parameter configurations.

Avenues for further work would include exploring the impact of more complex patterns of missingness in multiple predictor variables. As the number of incomplete predictors increases, so does the number of potential missing indicators eligible for inclusion in the outcome model, which could introduce issues of overfitting[26,27], and variable selection becomes challenging. Here we have also limited our studies to scenarios where the missingness mechanism remains constant between development and deployment, however it would be interesting to explore whether these same results hold if the mechanism were to change between the two stages.

## 5.1 Conclusion

We have conducted an extensive simulation study, and found that when no missingness is allowed at deployment, existing guidelines on how to impute missing data at development are generally appropriate. However, if missingness is to be allowed when the model is deployed into practice, the missing data handling strategy at each stage should be more carefully considered, and we found that omitting the outcome from the imputation model at all stages was optimal in terms of predictive performance.

We have found that RI performs at least as well as multiple imputation when the outcome is omitted from the imputation model, but tends to result in more unstable estimates of predictive performance. Missing indicators can offer marginal improvements in predictive performance under MAR and MNAR-X structures, but can harm model performance under MNAR-Y.

We recommend that if missing data is likely to occur both during the development and implementation of a CPM, that RI be considered as a more practical alternative to multiple imputation, and that if either imputation method is to be applied at deployment, the outcome be omitted from the imputation model at both stages. Model performance should be assessed in such a way that reflects how missing data will occur and be handled at deployment, as the most appropriate strategy may depend on whether missingness will be allowed once the model is applied in practice. We also advocate for the careful use of missing indicators in the outcome model if MNAR-X can safely be assumed, but this should be assessed on a study-by-study basis since the inclusion of missing indicators also has the potential to reduce predictive performance, especially when missingness is not permitted at deployment.

# 7 Supplementary Materials

## 7.1 Appendix 1: Calibration-in-the-large (calibration intercept)

We first present the computed performance metrics not included in the main manuscript: Calibration-in-the-large (calibration intercept) and the C-statistic.

The calibration intercept is defined as the intercept from a model fitted to the observed outcome with the linear predictor as an offset, and has a target value of 0. As in the main manuscript, the following parameters remain fixed throughout this section: $\gamma_{X1} = \gamma_{X2} = 0.7, \gamma_{X1X2} = 0.1, \pi_{R1} = 0.5$.

The results in are in line with the results of the other performance metrics - the calibration intercept is further away from 0 for models that use inconsistent imputation models between development and deployment. The inclusion of a missing indicator also appears to improve calibration when missingness is not allowed at deployment under MAR mechanisms, but can be harmful under MNAR-Y mechanisms.

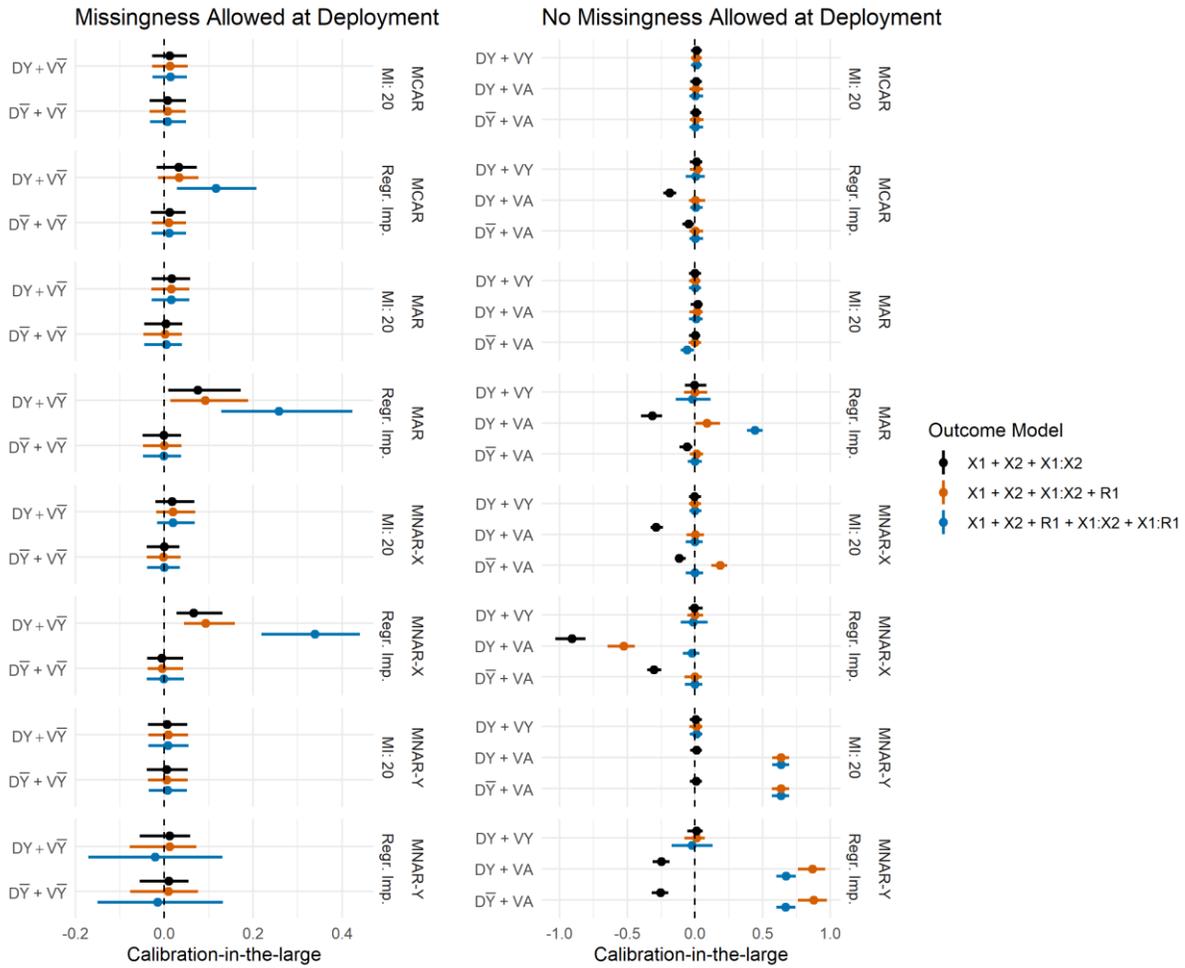

*Figure 7.1: Calibration-in-the-large estimates across imputation strategies, imputation methods and missingness mechanisms*

## 7.2 Appendix 2: C-statistic (calibration intercept)

shows the results for the C-statistic, using the same fixed parameter values as in the main text. Here we see that missing indicators can offer improvements in model discrimination under MNAR-Y (as expected) when used in combination with multiple imputation, but can cause some over-fitting when used in combination with regression imputation in this same context, as suggested by the calibration slopes that are slightly < 1 and the increased C-statistic from these models (MNAR-Y, regression imputation, missingness allowed at deployment).

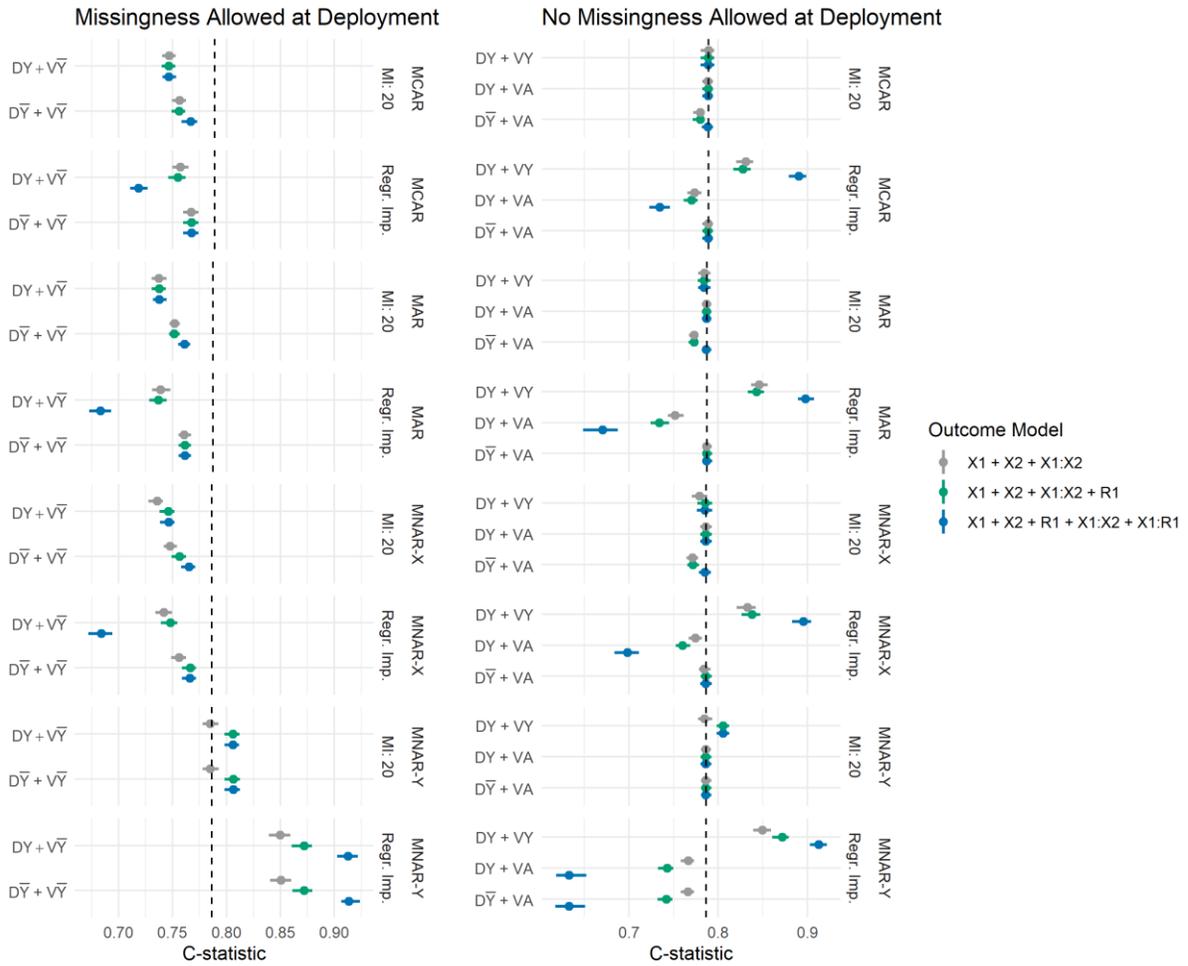

*Figure 7.2: C-statistic estimates across imputation strategies, imputation methods and missingness mechanisms*

## 7.3 Appendix 3: Predicted risk distribution

In  we present the distribution of the predicted risks, obtained from a single simulated dataset. The models were generated and validated omitting Y from the imputation model at both stages.

This was a post-hoc exploratory analysis, to explore the reason behind the seemingly optimistic performance estimates within this mechanism and missingness strategy. We can see from the plots that the predicted risks are quite extreme, with many very close to 0, and for the model containing the $X_1:R_1$ term, another peak towards 1. This would suggest that the model is overfit.

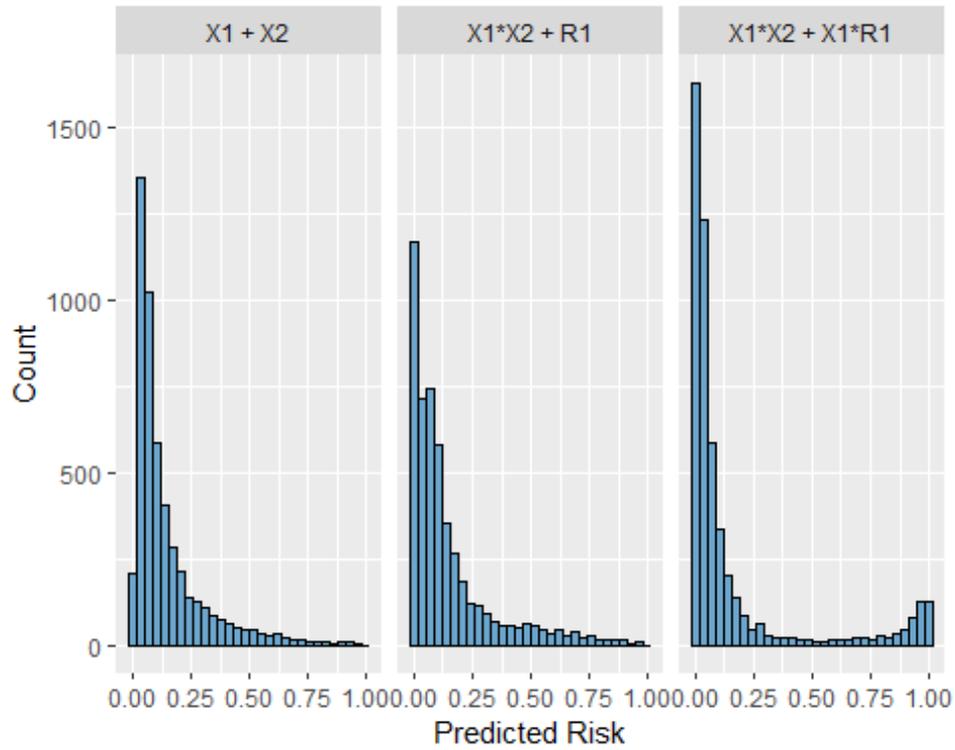

*Figure 7.3: Predicted risk distribution from model developed using regression imputation under MNAR-Y, missingness allowed and imputed at deployment, imputation model omitting the outcome at both stages*